\documentclass[aps,prl,twocolumn,superscriptaddress,showpacs]{revtex4}
\usepackage{graphicx}
\usepackage{dcolumn}
\newcommand{\zh}{z}

\newcommand{\xbj}{x}
\newcommand{\la}{\langle}
\newcommand{\ra}{\rangle}

\begin{document}

\title{{\large Measurement of Beam-Spin  Asymmetries for \\
$\pi^+$ Electroproduction Above the Baryon Resonance Region }}

\newcommand*{\JLAB }{ Thomas Jefferson National Accelerator Facility, Newport News, Virginia 23606} 
\affiliation{\JLAB } 

\newcommand*{\INFNFR }{ INFN, Laboratori Nazionali di Frascati, 00044 Frascati, Italy} 
\affiliation{\INFNFR }

\newcommand*{\ASU }{ Arizona State University, Tempe, Arizona 85287-1504} 
\affiliation{\ASU } 

\newcommand*{\SACLAY }{ CEA-Saclay, Service de Physique Nucleaire, F91191 Gif-sur-Yvette, Cedex, France} 
\affiliation{\SACLAY } 

\newcommand*{\UCLA }{ University of California at Los Angeles, Los Angeles, California  90095-1547} 
\affiliation{\UCLA } 

\newcommand*{\CMU }{ Carnegie Mellon University, Pittsburgh, Pennsylvania 15213} 
\affiliation{\CMU } 

\newcommand*{\CUA }{ Catholic University of America, Washington, D.C. 20064} 
\affiliation{\CUA } 

\newcommand*{\CNU }{ Christopher Newport University, Newport News, Virginia 23606} 
\affiliation{\CNU } 

\newcommand*{\UCONN }{ University of Connecticut, Storrs, Connecticut 06269} 
\affiliation{\UCONN } 

\newcommand*{\DUKE }{ Duke University, Durham, North Carolina 27708-0305} 
\affiliation{\DUKE } 

\newcommand*{\GBEDINBURGH }{ Edinburgh University, Edinburgh EH9 3JZ, United Kingdom} 
\affiliation{\GBEDINBURGH } 

\newcommand*{\FIU }{ Florida International University, Miami, Florida 33199} 
\affiliation{\FIU } 

\newcommand*{\FSU }{ Florida State University, Tallahassee, Florida 32306} 
\affiliation{\FSU } 

\newcommand*{\GWU }{ The George Washington University, Washington, DC 20052} 
\affiliation{\GWU } 

\newcommand*{\GBGLASGOW }{ University of Glasgow, Glasgow G12 8QQ, United Kingdom} 
\affiliation{\GBGLASGOW } 

\newcommand*{\INFNGE }{ INFN, Sezione di Genova, 16146 Genova, Italy} 
\affiliation{\INFNGE } 

\newcommand*{\ORSAY }{ Institut de Physique Nucleaire ORSAY, Orsay, France} 
\affiliation{\ORSAY } 

\newcommand*{\BONN }{ Institute f\"{u}r Strahlen und Kernphysik, Universit\"{a}t Bonn, Germany} 
\affiliation{\BONN } 

\newcommand*{\ITEP }{ Institute of Theoretical and Experimental Physics, Moscow, 117259, Russia} 
\affiliation{\ITEP } 

\newcommand*{\JMU }{ James Madison University, Harrisonburg, Virginia 22807} 
\affiliation{\JMU } 

\newcommand*{\KYUNGPOOK }{ Kyungpook National University, Taegu 702-701, South Korea} 
\affiliation{\KYUNGPOOK } 

\newcommand*{\MIT }{ Massachusetts Institute of Technology, Cambridge, Massachusetts  02139-4307} 
\affiliation{\MIT } 

\newcommand*{\UMASS }{ University of Massachusetts, Amherst, Massachusetts  01003} 
\affiliation{\UMASS } 

\newcommand*{\UNH }{ University of New Hampshire, Durham, New Hampshire 03824-3568} 
\affiliation{\UNH } 

\newcommand*{\NSU }{ Norfolk State University, Norfolk, Virginia 23504} 
\affiliation{\NSU } 

\newcommand*{\OHIOU }{ Ohio University, Athens, Ohio  45701} 
\affiliation{\OHIOU } 

\newcommand*{\ODU }{ Old Dominion University, Norfolk, Virginia 23529} 
\affiliation{\ODU } 

\newcommand*{\PITT }{ University of Pittsburgh, Pittsburgh, Pennsylvania 15260} 
\affiliation{\PITT } 

\newcommand*{\ROMA }{ Universita' di ROMA III, 00146 Roma, Italy} 
\affiliation{\ROMA } 

\newcommand*{\RPI }{ Rensselaer Polytechnic Institute, Troy, New York 12180-3590} 
\affiliation{\RPI } 

\newcommand*{\RICE }{ Rice University, Houston, Texas 77005-1892} 
\affiliation{\RICE } 

\newcommand*{\URICH }{ University of Richmond, Richmond, Virginia 23173} 
\affiliation{\URICH } 

\newcommand*{\SCAROLINA }{ University of South Carolina, Columbia, South Carolina 29208} 
\affiliation{\SCAROLINA } 

\newcommand*{\UTEP }{ University of Texas at El Paso, El Paso, Texas 79968} 
\affiliation{\UTEP } 

\newcommand*{\VT }{ Virginia Polytechnic Institute and State University, Blacksburg, Virginia   24061-0435} 
\affiliation{\VT } 

\newcommand*{\VIRGINIA }{ University of Virginia, Charlottesville, Virginia 22901} 
\affiliation{\VIRGINIA } 

\newcommand*{\WM }{ College of Willliam and Mary, Williamsburg, Virginia 23187-8795} 
\affiliation{\WM } 

\newcommand*{\YEREVAN }{ Yerevan Physics Institute, 375036 Yerevan, Armenia} 
\affiliation{\YEREVAN } 

\newcommand*{\NOWNCATU }{ North Carolina Agricultural and Technical State University, Greensboro, NC 27411}

\newcommand*{\NOWGBGLASGOW }{ University of Glasgow, Glasgow G12 8QQ, United Kingdom}

\newcommand*{\NOWJLAB }{ Thomas Jefferson National Accelerator Facility, Newport News, Virginia 23606}

\newcommand*{\NOWSCAROLINA }{ University of South Carolina, Columbia, South Carolina 29208}

\newcommand*{\NOWFIU }{ Florida International University, Miami, Florida 33199}

\newcommand*{\NOWINFNFR }{ INFN, Laboratori Nazionali di Frascati, Frascati, Italy}

\newcommand*{\NOWOHIOU }{ Ohio University, Athens, Ohio  45701}

\newcommand*{\NOWCMU }{ Carnegie Mellon University, Pittsburgh, Pennsylvania 15213}

\newcommand*{\INDSTRA }{ Systems Planning and Analysis, Alexandria, Virginia 22311}

\newcommand*{\NOWASU }{ Arizona State University, Tempe, Arizona 85287-1504}

\newcommand*{\NOWCISCO }{ Cisco, Washington, DC 20052}

\newcommand*{\NOWUK }{ Kentucky, LEXINGTON, KENTUCKY 40506}

\newcommand*{\NOWSACLAY }{ CEA-Saclay, Service de Physique Nucl\'eaire, F91191 Gif-sur-Yvette, Cedex, France}

\newcommand*{\NOWRPI }{ Rensselaer Polytechnic Institute, Troy, New York 12180-3590}

\newcommand*{\NOWUNCW }{ North Carolina}

\newcommand*{\NOWHAMPTON }{ Hampton University, Hampton, VA 23668}

\newcommand*{\NOWTulane }{ Tulane University, New Orleans, Lousiana  70118}

\newcommand*{\NOWKYUNGPOOK }{ Kyungpook National University, Taegu 702-701, South Korea}

\newcommand*{\NOWCUA }{ Catholic University of America, Washington, D.C. 20064}

\newcommand*{\NOWGEORGETOWN }{ Georgetown University, Washington, DC 20057}

\newcommand*{\NOWJMU }{ James Madison University, Harrisonburg, Virginia 22807}

\newcommand*{\NOWURICH }{ University of Richmond, Richmond, Virginia 23173}

\newcommand*{\NOWCALTECH }{ California Institute of Technology, Pasadena, California 91125}

\newcommand*{\MOSCOW }{ Moscow State University, General Nuclear Physics Institute, 119899 Moscow, Russia}

\newcommand*{\NOWVIRGINIA }{ University of Virginia, Charlottesville, Virginia 22901}

\newcommand*{\NOWYEREVAN }{ Yerevan Physics Institute, 375036 Yerevan, Armenia}

\newcommand*{\NOWRICE }{ Rice University, Houston, Texas 77005-1892}

\newcommand*{\NOWINFNGE }{ INFN, Sezione di Genova, 16146 Genova, Italy}

\newcommand*{\NOWBATES }{ MIT-Bates Linear Accelerator Center, Middleton, MA 01949}

\newcommand*{\NOWODU }{ Old Dominion University, Norfolk, Virginia 23529}

\newcommand*{\NOWVSU }{ Virginia State University, Petersburg,Virginia 23806}

\newcommand*{\UNIONC }{ Department of Physics, Schenectady, NY 12308}

\newcommand*{\NOWORST }{ Oregon State University, Corvallis, Oregon 97331-6507}

\newcommand*{\NOWCNU }{ Christopher Newport University, Newport News, Virginia 23606}

\newcommand*{\NOWGWU }{ The George Washington University, Washington, DC 20052}

  
\author{H.~Avakian}
     \affiliation{\JLAB}
     \affiliation{\INFNFR}
\author{V.D.~Burkert}
     \affiliation{\JLAB}
\author{L.~Elouadrhiri}
     \affiliation{\JLAB}
\author{N.~Bianchi}
     \affiliation{\INFNFR}
\author{G.~Adams}
     \affiliation{\RPI}
\author{A.~Afanasev}
     \affiliation{\JLAB}
\author{P.~Ambrozewicz}
     \affiliation{\FIU}
\author{E.~Anciant}
     \affiliation{\SACLAY}
\author{M.~Anghinolfi}
     \affiliation{\INFNGE}
\author{D.S.~Armstrong}
     \affiliation{\WM}
% ...
 
\collaboration{The CLAS Collaboration}
     \noaffiliation
%
 
%
%
\date{\today}
\begin{abstract}
We report the first evidence for a non-zero beam-spin azimuthal 
asymmetry in the
electroproduction of positive pions in the deep-inelastic kinematic region.
Data for the reaction $ep \rightarrow e'\pi^{+}X$ have been obtained using a polarized electron beam of 4.3 GeV
with the  CEBAF Large Acceptance 
Spectrometer  at the Thomas Jefferson National Accelerator
Facility (JLab). 
The amplitude
of the   $\sin\phi$ modulation increases with the 
momentum of the pion relative to the virtual photon, $z$.
In the range  $z=$0.5--0.8 the
average amplitude is  $0.038 \pm 0.005 \pm 0.003$ for a missing mass $M_X>$1.1 GeV and 
$0.037 \pm 0.007 \pm 0.004$ for $M_X>$1.4 GeV.

\end{abstract}
\maketitle

The origin of the spin of the proton has become a topic of considerable
experimental and theoretical interest since the EMC \cite{EMC} measurements  
implied that  quark helicities account for only a small 
fraction of the nucleon spin.
As a consequence, the study of the gluon polarization and the 
orbital angular momentum of partons  have become  
of central  interest. 
Single-Spin  Asymmetries (SSAs), measured in 
hadronic reactions for decades \cite{heller,E704}, 
have emerged as important observables to access transverse momentum
distributions of  partons and the
orbital angular momentum of quarks in the nucleon. 

In this paper we present the first measurement of a
non-zero beam-spin asymmetry in the electroproduction of
positive pions in deep-inelastic kinematic region.
Recently measurements of target SSAs have been reported 
for pion production  in semi-inclusive deep-inelastic scattering (SIDIS)
by the HERMES collaboration \cite{HER,HER2,HER4} 
for longitudinally polarized targets, and by the SMC collaboration
for a transversely polarized target \cite{SMC}.
Such SSAs require a correlation
between the spin direction of a particle and the orientation of the
 production (or scattering) plane, and have been  linked to the 
orbital angular momentum of partons in the nucleon \cite{BRODY,JIMA}.
Two fundamental QCD mechanisms giving rise to single spin asymmetries 
were identified. First the Collins mechanism \cite{COL,TM,KO}, where the
 asymmetry is generated
in the fragmentation of transversely polarized quarks, and second 
the Sivers
mechanism \cite{SIV,AM,BOMU,BRO,COLL,JIY,BELJI}, 
where it arises due to final state interactions
at the distribution function level.
The interference of  wavefunctions with different orbital angular
momentum, which is required to generate the SSA  \cite{BRO,COLL,JIY,BELJI}, also yields 
the helicity-flip Generalized Parton Distribution (GPD) 
$E$ \cite{BRODY,JIMA} that enters 
Deeply Virtual Compton Scattering (DVCS) \cite{Ji97,Rady} and the Pauli form 
factor $F_2$. 
The connection of SSAs and GPDs  has also been discussed in terms
of the transverse distribution of quarks in nucleons \cite{BURK,BURKHW}.

Physical observables accessible in SSAs include novel 
distribution and fragmentation functions such as chiral-odd transversity 
distribution
\cite{RS,JAF}, the  {\it time-reversal odd} (T-odd) distribution \cite{SIV,AM,BOMU,BRO,COLL,JIY,BELJI} and the  Collins \cite{COL} 
T-odd fragmentation functions.

In the partonic description
of SIDIS, distribution and fragmentation
functions  depend on the scaling variables $x$ and $z$,
respectively (see below for the definition). At fixed and moderate
values of the four momentum transfer $Q^2$ and at large
 values of $x$ and $z$, the contribution of multiparton
correlations or higher twist effects increases, eventually leading to a
breakdown of the partonic description. Model calculations indicate that 
SSAs are 
less sensitive to a number of higher order corrections
than cross section measurements in both semi-inclusive \cite{BAC} and 
in hard exclusive \cite{FRA,BEL} pion  production. The measurement of 
spin asymmetries could therefore become
a major tool for  studying  quark transverse momentum dependent distributions 
\cite{SIV,COL,TM,KO,BELJI,LM} and GPDs\cite{FRA,BEL}  in the 
$Q^2$ domain of a few {\rm GeV}$^2$. 

The beam-spin asymmetries
in single-pion production are higher-twist by their nature \cite{TM,AFCA,YUAN} and
are expected to increase at low $Q^2$.
Although large beam-spin asymmetries have been observed in measurements of
exclusive electroproduction of photons (DVCS) \cite{HERdvcs,CLASdvcs}, 
the only measurement of the 
beam-spin asymmetry in semi-inclusive pion 
electroproduction was 
reported recently by the HERMES collaboration \cite{HER}
at $\la z\ra \approx 0.4$. Within  statistical 
uncertainties their value
is consistent with zero.

The cross section for single pion production
by longitudinally polarized leptons 
scattering from unpolarized protons may be written in terms of a
set of response functions. 
The helicity 
($\lambda_e$) dependent part ($\sigma_{LU}$) 
\cite{TM,LM} arises from the 
anti-symmetric part of the hadronic tensor:

\begin{eqnarray}
\label{HLT}
\frac{d\sigma_{LU}}{d\xbj dy\,d\zh d^2P_\perp}
\propto \lambda_e\,\sqrt{y^2+\gamma^2}\sqrt{1-y-\frac{1}{4}\gamma^2}\,\sin \phi\,\,{\cal H}^\prime_{LT}.
\end{eqnarray}


\noindent The subscripts in $\sigma_{LU}$ specify the beam and  target 
polarizations, respectively ($L$ stands for longitudinally 
polarized and $U$ for unpolarized).
The  azimuthal angle $\phi$ is defined  by a triple product:

$$
\sin\phi = \frac{[\vec k_1 \times \vec k_2]\cdot\vec P_{\perp}}{|\vec k_1 \times \vec k_2|   |\vec P_{\perp}| },
$$

\noindent where
$\vec k_1$ and $\vec k_2$ are the initial and final electron momenta, and 
 $\vec P_{\perp}$ is the transverse 
momentum  of the  observed hadron with respect to
the virtual photon $\vec q$. 
The structure function ${\cal H}^\prime_{LT}$ arises due to the 
interference of the longitudinal and transverse photon contributions.
The kinematic variables $\xbj$, $y$, and $z$  are defined as: 
$
\xbj = Q^2/{2(P_1q)},\,\, y={(P_1q)/(P_1k_1)},\,\, \zh={(P_1P)/(P_1q)}, 
$
where $Q^2=-q^2$, $q=k_1-k_2$ is the 4-momentum 
of the virtual photon,
$P_1$ and $P$ are the momenta of the target and the observed final-state 
hadron, and $\gamma^2=4M^2\xbj^2y^2/Q^2$.

The beam-spin 
asymmetries  in single-pion semi-inclusive 
leptoproduction  were measured  using a 4.3 GeV
electron beam and the CEBAF Large Acceptance 
Spectrometer (CLAS) \cite{CLAS} at JLab.
Scattering of  longitudinally polarized electrons off
a liquid-hydrogen target was studied over a wide kinematic region.
The beam polarization, frequently measured  with a 
M{\o}ller polarimeter, was  on average $0.70 \pm 0.03$.
Beam helicity was flipped every 30 msec to minimize the helicity
correlated systematics.
The scattered electrons and pions were detected in
CLAS.
Electron candidates were selected by a hardware trigger using a 
coincidence between
the gas Cherenkov
counters and the lead-scintillator electromagnetic calorimeters. 
Pions
in a momentum range of 1.2 to 2.6 GeV were identified using momentum 
reconstruction in the tracking system
and the time-of-flight from the target to the timing scintillators. The total
number of electron-$\pi^+$ coincidences  
in the DIS range ($Q^2>1$ GeV$^2$, $W^2>4$ GeV$^2$)
was $\approx 4\times 10^5$.

A critical issue in SIDIS processes is the assumption of factorization,
i.e. that the hadron production cross section can be evaluated as
a convolution of a $x$-dependent distribution function, a hard scattering 
and a $z$-dependent fragmentation function. This picture is valid  
if the hadron originates from the fragmentation of the $current$
quark, assuming there is sufficient energy so that
the 4-momentum of the final hadron is not  directly related to 
that of the struck quark. 
At low $z$ hadrons may additionally originate from the fragmentation
of the target, while at large $z$, in addition to higher twist effects,
diffractive effects are also important.
Therefore a restricted range in $0.5<z<0.8$ has been selected for the
analysis.

\begin{figure}
\includegraphics[height=.28\textheight]{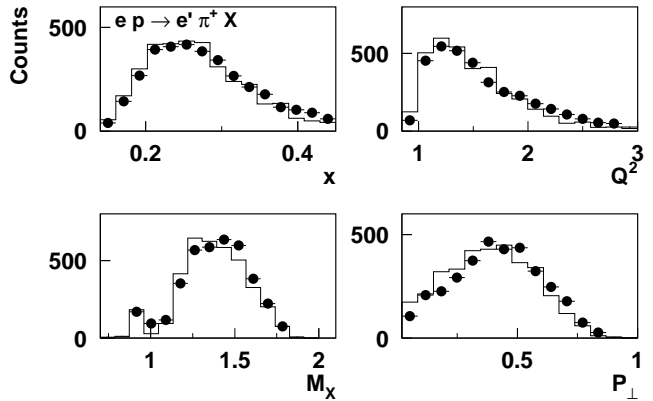}
\caption{ Comparison of the distributions measured with CLAS at 
 4.3 GeV  (circles) in $x,Q^2$(GeV$^2$), 
missing mass $M_X$(GeV), and the transverse pion momentum
 $P_{\perp}$(GeV) with LUND-MC reconstructed events. 
The distributions are averages over the range $0.5<z<0.8$; the 
MC results are normalized to the same number of events.
  }
 \label{fig:mismas}
 \end{figure}

The event distributions in the restricted $z$-range have  been compared with
a Monte Carlo based on the LUND generator \cite{pepsi} developed to describe
high energy processes. In LUND, the pion production
is dominated by direct production from string fragmentation, as opposed
to other processes such as target fragmentation and hadronic decays.
Figure \ref{fig:mismas} shows comparisons between the 
experimental yields
and the normalized MC distributions for different kinematical 
variables. The agreement of MC with the data suggests that the high 
energy-description of the SIDIS process can be extended to the
moderate energies of this measurement.

\begin{figure}
\includegraphics[height=.28\textheight]{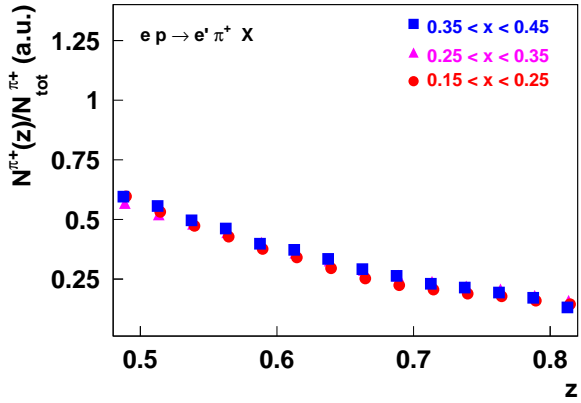}
\caption{Pion multiplicities as a function of $z$ for 
different $x$ ranges normalized by the total number 
of pions in the corresponding $x$-range. 
}
 \label{fig:zdist}
\end{figure}

To verify the factorization ansatz, pion multiplicities
have been extracted for different $x$ ranges. 
Pion multiplicities have been shown
to be approximately equivalent to fragmentation functions
\cite{PASQ}  which
depend on the $z$ variable only at fixed $Q^2$.
In Fig. \ref{fig:zdist} $\pi^+$ multiplicities normalized by 
the total number of events are shown as a function
of $z$ for different $x$ bins.  Within the range and the precision
of the present measurement no $x$-dependence of multiplicities is
seen. This experimental finding  
is also consistent with the assumption of factorization.

\begin{figure}
      \centerline{\includegraphics[height=.28\textheight]{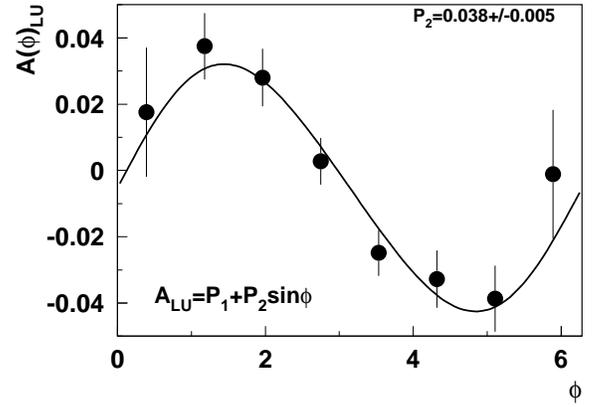}}
\caption{ The beam-spin azimuthal asymmetry  as a function of 
azimuthal angle $\phi$, 
measured in the range $z$=0.5--0.8.
  }
 \label{fig:sol}
 \end{figure}

The azimuthal distribution of the beam-spin
asymmetry for $\pi^+$:
 \begin{equation}
	\label{asy}
   A(\phi)_{LU}=\frac{1}{P}\frac{N^+ - N^-}{N^+ + N^-},
 \end{equation}
is shown in  Fig. \ref{fig:sol}.
Here  $N^\pm$   
is the number of events for positive/negative
helicities of the electron and $P$ is the beam polarization.
The data show a clear $\sin\phi$ modulation
from which a $\sin\phi$  moment of $0.038\pm0.005(stat)$ can be
derived.

The same quantity can be
formed by
extracting moments of the cross section for the two 
helicity states weighted by 
the corresponding $ \phi$-dependent functions.
In this case the $\sin\phi$ moment is given by:
\begin{equation}
\label{sinfi}
A_{LU}^{\sin\phi}=
\frac{2}{P N^\pm}\sum_{i=1}^{N^\pm} \sin\phi_i,
\end{equation}

\begin{figure}
      \centerline{\includegraphics[height=.28\textheight]{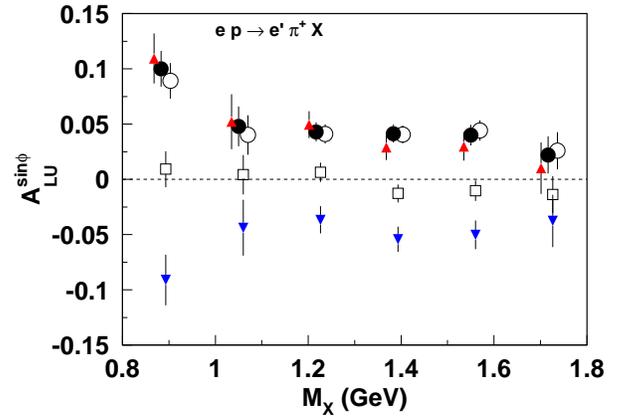}}
\caption{ The beam-spin azimuthal asymmetry  as a function of 
missing mass 
$M_X$, 
in $\gamma^*p\rightarrow \pi^+X$ 
extracted in the range $0.5<z<0.8$. 
Triangles up and down 
are the results for positive and negative helicities, respectively, and 
the filled circles are for their average.
Open circles show the measured $A_{LU}^{\sin\phi}$ extracted as a 
$\sin \phi$ moment of the spin asymmetry. Open squares show the 
measured $A_{LU}^{\sin\phi}$ for the sample averaged over the
beam polarization. Data are slightly shifted in $M_X$ for clarity.
  }
 \label{fig:alumismas}
 \end{figure}

The two methods are identical for a full acceptance detector, 
but in practice 
have different sensitivities to acceptance effects.
In Fig. \ref{fig:alumismas} the comparison of ${\it A}^{\sin\phi}_{LU}$
derived with the two methods is presented as a function of 
the missing mass $M_X$ evaluated in the $e p \rightarrow e' \pi^+ X$
reaction. As can be seen the results
agree well with each other over the full $M_X$ range.
 The moments  $A_{LU}^{\sin\phi}$
can be also computed for each helicity state and for an average 
of zero helicity which corresponds to an unpolarized beam. These data
shown in Fig. \ref{fig:alumismas} provide an additional
test of the absence of spurious azimuthal asymmetries.
All these results indicate that within the statistical
uncertainties, the acceptance corrections are small and
under control.

Contributions to the systematic uncertainties arise
from spin-dependent moments 
of the cross section coupling
to corresponding moments in the acceptance to produce 
corrections
to the measured  $\sin \phi$ moment \cite{HER}. 
 The contribution to uncertainties due to 
the CLAS acceptance in all relevant kinematic
variables ($x,y,z,P_\perp,\phi$)
is evaluated to be less than 0.004 in average
and less than 0.007 in all bins.
The systematic uncertainties in the measurement of the beam polarization 
contribute at an even lower level (0.002). 
Possible particle mis-identification over the accessible kinematic range changes the observed
SSA by less than 0.001. 
To minimize radiative corrections, a cut on the energy 
of the virtual photon relative to the incoming
electron ($y<0.85$) was imposed.
The estimated radiative corrections 
do not exceed a few percent of the value
of the SSA \cite{AFA}, and  give a minor contribution to the systematic uncertainty.
Other systematic uncertainties are negligible.

As can be seen in Fig. \ref{fig:alumismas}, the missing mass on the remnant
system is mostly occurring in the nucleon resonance region.
Despite this, the beam SSA doesn't show a dependence on any
specific final state. A sizable increase of the SSA only
appears in  exclusive $\pi^+$ production
where the missing mass corresponds to the nucleon mass.
For this reason, two different cuts on the missing mass $M_X$ have been applied 
in the final analysis. A first cut at $M_X > 1.1$ GeV was chosen to exclude
the contribution of the exclusive $\pi^+$ production on the nucleon.
A higher cut at $M_X > 1.4$ GeV was also considered to reduce, in
addition, the contribution of semi-exclusive $\pi^+$ production with a recoiling
$\Delta^0$-resonance.
For $M_X>$  1.4 GeV multi-hadron production is the dominant mechanism
and the possible contribution of higher nucleon resonances in the
recoiling system can be interpreted in terms of quark-hadron duality.
This seems to be supported by the smooth and almost flat behavior of 
the data shown in Fig. \ref{fig:alumismas}.
 
\begin{figure}
      \centerline{\includegraphics[height=.28\textheight]{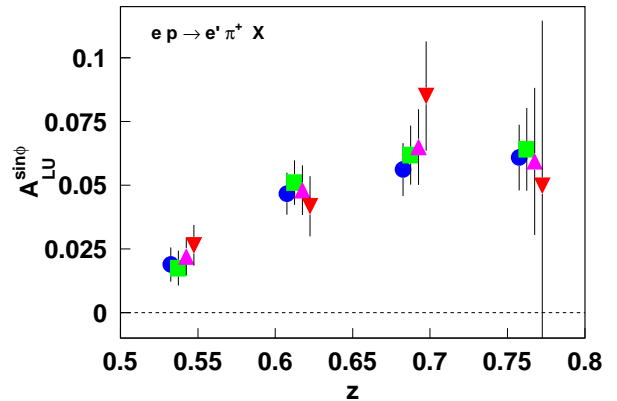}}
\caption{ The beam-spin azimuthal asymmetry  as a function of $z$ extracted for
different cuts on the missing mass 
$M_X$ (in GeV), $M_X > 1.1$ (circles), $M_X > 1.2$ (squares),  $M_X > 1.3$ (triangles up) and  $M_X > 1.4$ (triangles down).} 
 \label{fig:e1c_zdepcomp}
 \end{figure}

A further check is shown in Fig. \ref{fig:e1c_zdepcomp},
 where the $z$-dependence
of the beam SSA is presented for increasing values of
the missing mass cut. Due to the large correlation between
the $z$ and the $M_X$ variables, an increasing $M_X$ cut
drastically reduces the number of events with large $z$,
leaving almost unchanged the beam SSA.
This indicates that, within the present statistical
uncertainties, the fraction $z$ of the virtual photon energy
carried by the pion, rather than the missing mass $M_X$, 
is the relevant variable in the scattering process.

\begin{figure}
\includegraphics[height=.28\textheight]{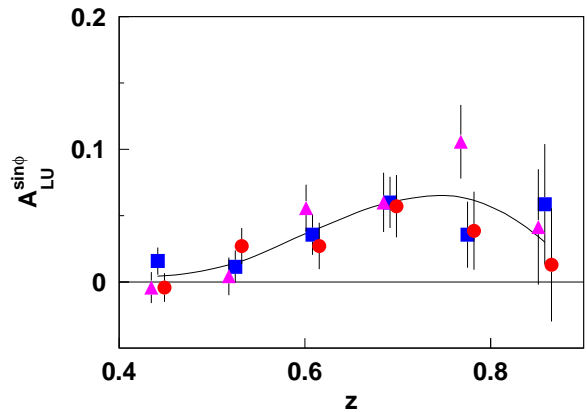}
\caption{Beam SSA as a function of  $z$ for 
different $x$ ranges (same as in Fig. \ref{fig:zdist}) 
for $M_X>1.1$ GeV. The curve is a simple fit to all data to show
the general behavior of the asymmetry. }
 \label{fig:zalu}
\end{figure}

Consistency with the factorization assumption, which has been
already shown in Fig. \ref{fig:zdist} for $\pi^+$ multiplicities, 
has also been
investigated for the observable under study. In Fig. \ref{fig:zalu}
the beam SSA is presented as a function of $z$ for different $x$-bins.
Its general behavior suggested by the curve  
does not exhibit any significant $x$-dependence, which 
 is also consistent with 
factorization in the chosen kinematic range.

\begin{figure}
\includegraphics[height=.28\textheight]{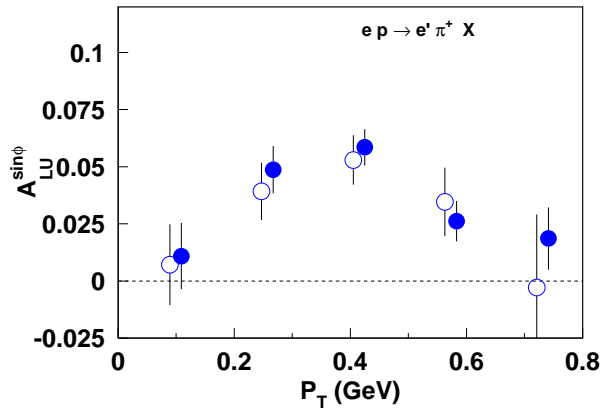}
\caption{Beam SSA as a function of  $P_{\perp}$ 
for $M_X>1.1$ GeV (filled circles) and $M_X>1.4$ GeV (open circles).} 
 \label{fig:ptdist}
\end{figure}

The dependence of beam SSA on the $\pi^+$ transverse momentum, $P_\perp$,  
is shown in  Fig. \ref{fig:ptdist}.
A linear dependence of SSA 
 on the  $P_\perp$ (with kinematic zero at $P_\perp=0$ GeV) is expected,
at least for the moderate range of $P_\perp$ \cite{TM,KO}.

\begin{figure}[h]
    \centerline{\includegraphics[height=.28\textheight]{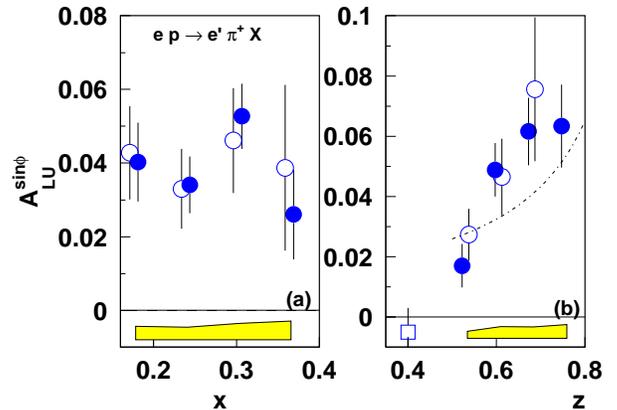}}
\caption{ The beam-spin azimuthal asymmetry  as a function of $\xbj$ (a),
in the range   $0.5<z<0.8$, and as function of $z$ (b), in the
 range $0.15<\xbj<0.4$ for $M_X>1.1$ GeV (filled circles).
 The error bars show the
statistical uncertainty, and the band 
represents the systematic uncertainties. The empty  circles are results 
for $M_X>1.4$ and have been slightly shifted to make them more visible. 
The empty  square 
shows the HERMES result \cite{HER}, which is an average  over the
range $z$=0.2--0.7 and $x$=0.02--0.4. The curve is  a theoretical
prediction \cite{YUAN}. }
 \label{fig:e1c_xzdep}
 \end{figure}

The beam spin asymmetries averaged over the two spin states
as a function of $x$ and $z$ are plotted in Fig.~\ref{fig:e1c_xzdep}
and listed in Table  \ref{tab:xdata}.
Table \ref{tab:xkin} shows the relevant variables for the
different $x$ and $z$ bins. 
The beam SSA is positive for a positive electron helicity
over the entire measured range.
The data do not show a significant $x$-dependence while the asymmetry
is strongly increasing at high $z$, where according to LUND MC,
the probability of the detected pion to carry the struck quark
is maximal.

The size and the behavior of the asymmetry are very similar for the 
two cases of missing mass cuts. 
For the case of higher $M_X$ cut, the data are compared with a
prediction \cite{YUAN} based on the Sivers mechanism as the dynamical
origin of the observable. Within this framework, the asymmetry
is given by the convolution of the T-odd parton distribution
$h_1^{\perp}$ with the twist-3 fragmentation function $E(x)$ \cite{JAF1,JI2}.
The latter function is responsible for the strong $z$-dependence
of the asymmetry. Despite the fact that the formalism is much better suited
for higher energy reactions, the agreement in size and behavior
with the data is reasonable.

The CLAS preliminary data on the beam SSA have been also 
interpreted in terms of the Collins mechanism 
\cite{LM,EFRPI1,EFRhep,GAM} for a first
determination of the twist-3 distribution function
$e(x)$ \cite{EFRPI1,EFRhep}. 
The magnitude of their $e(x)$ is also consistent with
predictions using the chiral quark soliton model \cite{SHW1,WAK1,WAK2}. 

At lower $z$ HERMES reported results 
consistent with zero beam-spin asymmetry \cite{HER}.
CLAS asymmetries, obtained at higher $z$ and lower $Q^2$,
should be kinematically enhanced with
respect to the higher energy HERMES data \cite{HER}, 
as pointed out in Refs. \cite{AFCA,YUAN},
so that no evident contradiction can be inferred between the
two measurements.

The data at lower $M_X$ demonstrate that for this observable
the transition from the semi-inclusive to the semi-exclusive and to
the exclusive domains is smooth.
In addition these data may provide a new field for testing, 
in the final state, the quark-hadron duality, which has been proved
to work in the initial state for other observables,
like the spin-independent \cite{Nicul} and the spin-dependent
\cite{Airap} structure functions, down to low values of $Q^2$.

In conclusion,
we have presented the first measurement of a non-zero  beam-spin asymmetry
in single  $\pi^+$ inclusive electroproduction in the DIS kinematic region. 
The average asymmetry is $0.038 \pm 0.005 \pm 0.003$ for a missing mass $M_X>$1.1 GeV and 
$0.037 \pm 0.007 \pm 0.004$ for $M_X>$1.4 GeV.
The asymmetry shows a strong enhancement at large values of $z$
while no significant $x$-dependence is present within the measured range.
Detailed experimental and Monte Carlo studies have been performed
showing no large violation of the factorization assumption for
the process and suggesting that the partonic description may
be applied in the kinematical range of the measurement.
New data are expected from experiments utilizing a 6 GeV polarized beam,
allowing a more precise test of the factorization ansatz and
the investigation of the $Q^2$ dependence of the
asymmetries.

We thank S. Brodsky, M. Diehl, A. Efremov, A. Kotzinian, A. Metz, and  P. Mulders 
for stimulating discussions.
We would like to acknowledge the outstanding efforts of the staff of the 
Accelerator and the Physics Divisions at JLab that made this experiment possible.
This work was supported in part by the U.S. Department of Energy
and the National  Science Foundation, 
the Italian Istituto Nazionale di Fisica Nucleare, the 
 French Centre National de la Recherche Scientifique, 
the French Commissariat \`{a} l'Energie Atomique, 
an Emmy Noether grant from the Deutsche 
Forschungsgemeinschaft, and the Korean Science and Engineering Foundation.
The Southeastern Universities Research Association (SURA) operates the 
Thomas Jefferson National Accelerator Facility for the United States 
Department of Energy under contract DE-AC05-84ER40150. 

\vspace{-0.22cm}

\vspace{1.3cm}

\begin{table}[ht]
\caption{\label{tab:xdata} SSA: $\xbj$ and $z$-dependence for $M_X>1.1$ GeV (upper table) and $M_X>1.4$ GeV (lower table).}
\begin{tabular}{l|l|c|c|r|r|}
\hline
 $ x $    & $A_{LU}^{\sin\phi} \pm \Delta_{stat} \pm \Delta_{syst}$ & &$  z $  & $A_{LU}^{\sin\phi} \pm \Delta_{stat} \pm \Delta_{syst}$\\
\hline
0.18    & 0.041   $\pm$ 0.011 $\pm$ 0.004 && 0.54    & 0.017   $\pm$ 0.007  $\pm$ 0.002 \\
0.24    & 0.034   $\pm$ 0.008 $\pm$ 0.003 && 0.61    & 0.049   $\pm$ 0.009  $\pm$ 0.004  \\
0.31    & 0.053   $\pm$ 0.009 $\pm$ 0.004 && 0.69    & 0.062   $\pm$ 0.011  $\pm$ 0.004\\
0.37    & 0.026   $\pm$ 0.012 $\pm$ 0.005 && 0.76    & 0.063   $\pm$ 0.014  $\pm$ 0.005 \\
\hline
\hline
0.18    & 0.043   $\pm$ 0.013 $\pm$ 0.005 && 0.54    & 0.027   $\pm$ 0.009  $\pm$ 0.003 \\
0.24    & 0.033   $\pm$ 0.011 $\pm$ 0.004 && 0.61    & 0.047   $\pm$ 0.013  $\pm$ 0.005  \\
0.31    & 0.046   $\pm$ 0.014 $\pm$ 0.005 && 0.69    & 0.076   $\pm$ 0.024  $\pm$ 0.005\\
0.37    & 0.039   $\pm$ 0.023 $\pm$ 0.007 && 0.76    & 0.067   $\pm$ 0.080  $\pm$ 0.007 \\
\hline
\end{tabular}
\end{table}

\begin{table}[ht]
\caption{Average values of $Q^2$ (GeV$^2$), $W$ (GeV), $y$,  $P_{\perp}$ (GeV) and   $z$/$\xbj$
 in each of $\xbj$ and $z$ bins for $M_X>1.1$ GeV (upper table) and $M_X>1.4$ GeV (lower table).}
 \label{tab:xkin}
\begin{tabular}{l|l|l|l|l|l|l|c|l|l|l|l}
\hline
$x$ & $\la Q^2 \ra$ & $\la W \ra$ & $\la y \ra$  & $\la P_{\perp} \ra$ &  $\la z \ra$        &&   $z$  & $\la Q^2 \ra$ & $\la W \ra$ & $\la P_{\perp} \ra$ & $\la x \ra$  \\
\hline
0.18    & 1.1    & 2.5 & 0.75 &  0.46  & 0.61 &&  0.54     & 1.46    & 2.3  &  0.43  &  0.27 \\
0.24    & 1.3    & 2.3 & 0.67 &  0.42  & 0.61 &&  0.61     & 1.44    & 2.3  &  0.43  &  0.27 \\
0.31    & 1.6    & 2.2 & 0.66 &  0.41  & 0.61 &&  0.69     & 1.44    & 2.3  &  0.42  &  0.27  \\
0.37    & 2.0    & 2.2 & 0.67 &  0.39  & 0.61 &&  0.77     & 1.43    & 2.3  &  0.36  &  0.27 \\
\hline
\hline
0.18    & 1.1    & 2.5 & 0.76 &  0.43  & 0.58&&  0.54     & 1.44    & 2.3 &  0.38  &  0.26 \\
0.24    & 1.4    & 2.3 & 0.70 &  0.36  & 0.57&&  0.61     & 1.41    & 2.4 &  0.34  &  0.25 \\
0.31    & 1.7    & 2.3 & 0.69 &  0.32  & 0.56&&  0.69     & 1.37    & 2.4 &  0.30  &  0.23  \\
0.37    & 2.0    & 2.2 & 0.70 &  0.28  & 0.56&&  0.77     & 1.26    & 2.5 &  0.24  &  0.20 \\
\hline
\end{tabular}
\end{table}

\end{document}